\documentclass[manuscript]{acmart}
\AtBeginDocument{%
  \providecommand\BibTeX{{%
    \normalfont B\kern-0.5em{\scshape i\kern-0.25em b}\kern-0.8em\TeX}}}

\copyrightyear{2024}
\acmYear{2024}
\setcopyright{rightsretained}
\acmConference[ASSETS '24]{The 26th International ACM SIGACCESS Conference on Computers and Accessibility}{October 27--30, 2024}{St. John's, NL, Canada}
\acmBooktitle{The 26th International ACM SIGACCESS Conference on Computers and Accessibility (ASSETS '24), October 27--30, 2024, St. John's, NL, Canada}
\acmDOI{10.1145/3663548.3688501}
\acmISBN{979-8-4007-0677-6/24/10}

\usepackage{multirow}
\usepackage{subfig}
\begin{document}


\title[Predictive Anchoring]{Predictive Anchoring: A Novel Interaction to Support Contextualized Suggestions for Grid Displays}

\author{Cynthia Zastudil}
\email{cynthia.zastudil@temple.edu}
\affiliation{%
 \institution{Temple University}
 \streetaddress{1801 N. Broad St.}
 \city{Philadelphia}
 \state{Pennsylvania}
 \country{USA}
 \postcode{19122}
}

\author{Christine Holyfield}
\email{ceholyfi@uark.edu}
\affiliation{%
\institution{University of Arkansas}
\streetaddress{1 University of Arkansas}
\city{Fayetteville}
\state{Arkansas}
\country{USA}
\postcode{72701}}

\author{June A. Smith}
\email{smithj7@berea.edu}
\affiliation{%
\institution{Berea College}
\streetaddress{101 Chestnut St.}
\city{Berea}
\state{Kentucky}
\country{USA}
\postcode{40403}}

\author{Hannah Vy Nguyen}
\email{hannahvyng@gmail.com}
\affiliation{
\institution{Temple University}
\streetaddress{Computer and Information Sciences}
\city{Philadelphia}
\state{Pennsylvania}
\country{USA}
\postcode{19122}
}

\author{Stephen MacNeil}
\email{stephen.macneil@temple.edu}
\affiliation{%
 \institution{Temple University}
 \streetaddress{1801 N. Broad St.}
 \city{Philadelphia}
 \state{Pennsylvania}
 \country{USA}
 \postcode{19122}
}

\renewcommand{\shortauthors}{Zastudil et al.}

\begin{abstract}
Grid displays are the most common form of augmentative and alternative communication device recommended by speech-language pathologists for children. Grid displays present a large variety of vocabulary which can be beneficial for a users' language development. However, the extensive navigation and cognitive overhead required of users of grid displays can negatively impact users' ability to actively participate in social interactions, which is an important factor of their language development. We present a novel interaction technique for grid displays, Predictive Anchoring, based on user interaction theory and language development theory. Our design is informed by existing literature in AAC research, presented in the form of a set of design goals and a preliminary design sketch. Future work in user studies and interaction design are also discussed.
\end{abstract}




\received{3 July 2024}
\received[Accepted]{9 August 2024}

\maketitle

\section{Introduction}

Millions of children in the United States alone~\cite{shriberg_prevalence_1999, law_prevalence_2000} are on the autism spectrum\footnote{For this paper, we will be using identity-first language; however, it should be noted this is preferred by some autistic individuals, but not all individuals who identify as autistic~\cite{sharif2022should, taboas2023preferences, dwyer2022first}.}
or have other developmental disabilities who do not have functional speech and would therefore benefit from augmentative and alternative communication (AAC) devices. AAC devices provide children who do not have functional speech with a means of effective communication and support their learning and use of language~\cite{branson_use_2009, light2019new, woods_early_2003}. 

\begin{figure*}
\centering
\includegraphics[width=0.7\linewidth]{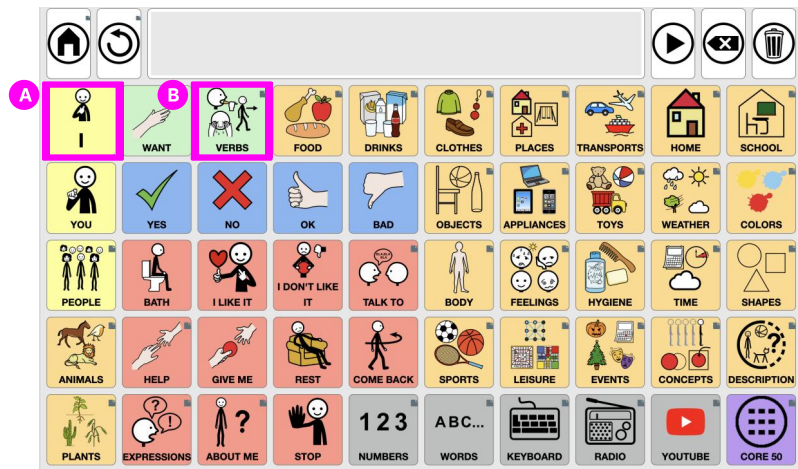}
\caption{An example grid display (source: \href{https://grid.asterics.eu/}{https://grid.asterics.eu/}). Users can select vocabulary using buttons with pictographs depicting the associated vocabulary (A) or navigate to other vocabulary using folders containing sub-pages of vocabulary (B).}
\Description[An example grid display]{This is an image depicting a grid display that an AAC user may use. There is a mix of buttons with different functionality. One kind of button is a communication option button which populates the text box if clicked. This highlighted example in this image is a button representing the word ``I'' which also contains a stick drawing representation of the word (the stick figure pointing to itself). The other kind of button is a folder button which allows the user to navigate to sub-pages of the application to access more vocabulary. The highlighted example in this image is of the folder for ``verbs'' which contains pictographs of various verbs (e.g., crying, eating, and walking).}
\label{fig:grid-example}
\end{figure*}

Grid displays are one of the most common forms of AAC for children~\cite{beukelman2020augmentative, thistle2015building}, in which communication options, usually containing pictograph representations of words, are arranged into a row-column format. These buttons are often grouped by function or topic into different pages within the grid display. A sample grid display is shown in Figure~\ref{fig:grid-example}. Users can select buttons to construct output which is often played via text-to-speech functionality~\cite{light2019designing_effective}. 

Grid displays are effective in supporting communication skill development, communicating users requests, and facilitating engagement~\cite{walker2013effects, romski2010randomized, ganz2012meta}. While grid displays can enhance communication outcomes, many AAC users and their communication partners are dissatisfied with the learnability and slow rate of communication in current AAC systems~\cite{rackensperger2005first, mcnaughton2008child}. These limitations hinder effective social interactions and make it difficult for users to fully engage in conversations~\cite{ostvik2018interactional, Batorwicz2014social, anderson2011he}.
These barriers to active social interactions are especially prevalent with younger, beginner communicators who are just starting to develop both their linguistic abilities and their skills using their AAC device~\cite{smith2009effect, light2002there}. Social interaction is a vital part of early language development~\cite{hoff2006social, vygotsky2012thought} and the inability to effectively communicate with others can cause frustration and emotional distress~\cite{mccormack2010my, mirenda1997supporting, walker2013effects}. There is an urgent need to explore how grid displays can be enhanced for beginning communicators to support more active participation in social interactions, and also to provide better opportunities to learn language~\cite{light2019new}.

In this poster, we leverage prior research from interaction design and linguistic development to propose Predictive Anchoring---a technique to enhance grid displays by anchoring vocabulary suggestions to communication options. The potential benefits include: 

\begin{itemize}
    \item Enhancing users' rate of communication by suggesting relevant vocabulary faster
    \item Supporting linguistic development by modeling language though contextualized vocabulary expansions as opposed to reorganizing the grid or presenting suggestions in a decontextualized space within the grid display.
    \item Supporting existing linguistic skills of the target user population by providing more accessible communication options which are relevant to the context they are interested in
\end{itemize}

We also identify three design goals that have been informed by prior work and open research areas within AAC, which may inform future work in designing enhanced grid displays.

\section{Related Work}

\subsection{Grid Displays}

AAC devices vary on a number of features, one of which being layout type~\cite{beukelman2020augmentative}. Most commonly, AAC layout follows a grid format in which visual symbols (e.g., pictographs) are displayed in a row-column format to represent words with corresponding voice output upon selection. Grid displays allow for a large number of words to be available for communication at once. This layout reduces navigational demands for individuals who use AAC who have large vocabularies and need to express a large number of words efficiently~\cite{gevarter2020dynamic, light_performance_2004}. However, grid displays introduce their own demands that can limit the accessibility of AAC technology for beginning communicators including visual-cognitive processing demands due to visual clutter~\cite{light2019designing_effective, wilkinson2022judicious} and linguistic demands due to the removal and isolation of words from context~\cite{light1997let}. While grid displays are not ideal for many beginning communicators~\cite{light2019new, light2007aac, shane2006using}, they continue to dominate the layouts featured by most available AAC technologies and they continue to be the most frequently selected layout by professionals such as SLPs~\cite{thistle2015building}. 


\subsection{Just-in-Time Programming for AAC Devices}

Just-in-Time (JIT) programming in an AAC context refers to the process followed by both AAC users and their communication partners in configuring AAC devices as scenarios for AAC use occur~\cite{holyfield_programming_2019, schlosser2016just}. This process is typically done manually by a communication partner (e.g., a SLP), but it can also be done automatically by leveraging a combination of users' context and machine learning algorithms~\cite{kane2012what, obiorah2021designing, devargas2022automated, Valenica2024Compa,valencia2023less,devargas2024codesigning}. Recently, numerous researchers are studying JIT programming to improve the rate of communication in both grid displays and VSDs~\cite{obiorah2021designing, Mooney_Bedrick_Noethe_Spaulding_Fried-Oken_2018, valencia2023less, Valenica2024Compa, kane2012what, holyfield2024leveraging, curtis2022state}. Numerous approaches attempt to enhance grid displays ~\cite{devargas2022automated, devargas2024codesigning} by using context in the form of photographs to automatically configure a grid display for the user. 
However, these approaches often rely on text prediction~\cite{garcia2015measuring, schadle2004sibyl}, which place text suggestions away from the words being selected and force split attention between choosing words and attending to the suggestions, or rearranging the grid display when accounting for contextual information~\cite{patel2007enhancing, garcia2016evaluating}, which leads to users having to visually process all of the grid options repeatedly, like keys on a keyboard moving around based on each keypress. This kinds of predictions may negatively impact users' ability to quickly search for and select communication options~\cite{light2019designing_effective}. 


\section{Design of Predictive Anchoring}

\subsection{Design Goals}


Grid displays typically require extensive and time-consuming navigation which negatively impacts users' ability to actively participate in social interactions~\cite{higginbotham2007access, johnston2004supporting}. When communications are timely, they align with the immediate context, ensuring that AAC users can actively participate in conversations and respond appropriately. This relevance helps maintain the interest and engagement of communication partners, preventing frustration and social isolation for AAC users. This is especially important for beginning communicators who often interact with other children who may have shorter attention spans~\cite{allyson2023experience, Batorwicz2014social}. \textbf{Design Goal 1: Quickly provide just-in-time contextually relevant communication options in a grid display.}


Learning how to use new AAC devices is an immense challenge~\cite{mcnaughton2008child, rackensperger2005first, allyson2023experience}. Given the prevalence of grid displays~\cite{thistle2015building} it is important to build on this system rather than require users to adopt and learn new tools.
As such, we do not want to propose the adoption of an entirely new system with a completely new method of interaction. Additionally, the existing organization of communication options within the grid display should remain unchanged to respect the mental models users construct while using the AAC device~\cite{wilkison2004contributions, cafiero2007aac, stuart2008case}. \textbf{Design Goal 2: We need to respect the existing interfaces that AAC users are accustomed to using.}


In research conducted on visual scene displays, another form of AAC, researchers have found that explicitly anchoring communication options to their real-world referents is beneficial for communication outcomes~\cite{drager_aac_2019, holyfield_effect_2019, o2016brief, light2012effects}. Grid displays traditionally present vocabulary independent of context. This imposes significant linguistic demands on users, especially those who are learning to use the AAC system~\cite{light1997let}. There is potential for exploring how vocabulary can be presented in a manner which employs the common grid display systems and preserves context by anchoring more communication options to icons within a grid display. \textbf{Design Goal 3: Vocabulary suggestions should be anchored to user interactions with other communication options.}

\subsection{Theoretical Underpinnings that Inform Predictive Anchoring}
Users face two primary problems with grid displays: 1) there are too many options to consider~\cite{light2012supporting_} and 2) techniques to automatically suggest content either decontextualize the suggestions by placing them far from the selected word~\cite{Pereira2022pictobert, garcia2016evaluating} or rearrange the words which can be confusing~\cite{devargas2022automated, devargas2024codesigning}. 

Predictive Anchoring leverages existing theories in user interface design and language development theory in order to begin iteration on a design which is based on existing evidence. A common user interaction strategy leveraged by designers is Hick's Law~\cite{hick1952rate}. Hick's Law states that the amount of time it takes for users to make decisions increases with the number and complexity of choices that are presented to users. Common strategies designers may use to reduce the amount of time it takes for users to make decisions are: 1) reduce the number of choices when quick decision-making is important, 2) breaking complex tasks down into simpler tasks (e.g., reducing the amount of navigation required), and 3) suggesting recommended options~\cite{liu2020how}. 
To implement these design guidelines per Hick's Law, our design enables users to highlight vocabulary of interest, then provides them with a subset of suggested communication options which may also be of interest to them.

Predictive Anchoring is not only supported by existing user interface techniques, but also by language development theory. First, children who are beginning communicators use conversation that is mostly relevant to their current environment~\cite{light1997let}. That is, they use words that they see in front of them and that are of high relevance to the social interaction. By anchoring communication options to just-used vocabulary, this approach aligns with beginning communicators' language strengths and offers communication options more likely to be meaningful to them~\cite{holyfield_programming_2019}. Furthermore, beginning communicators acquire more advanced language and communications skills by having the opportunity to use new words within social interactions~\cite{kaiser2011advances}. Beginning communicators will have the opportunity to communicate more words within an interaction by having quick access to high relevance, anchored communication options. Finally, as beginning communicators' language grows, they add words to single words they use often, representing both a widening of vocabulary and a heightened complexity of language output~\cite{wetherby1988analysis}. To support this language growth, parents and professionals such as SLPs often use a communication strategy called ``expansion'' in which they repeat the word communicated by the child learning language in a longer utterance by adding one or more additional words~\cite{paul1997facilitating}. For example, a child might point to a plane in the sky and say ``plane''. In response, a parent is likely to expand on the child's utterance, for instance by saying ``a plane flying''. Anchored communication options from the AAC technology, therefore, reflect good partner strategies that are supportive of language growth.

\subsection{Envisioned System Design}

\begin{figure*}
\centering
\includegraphics[width=0.8\linewidth]{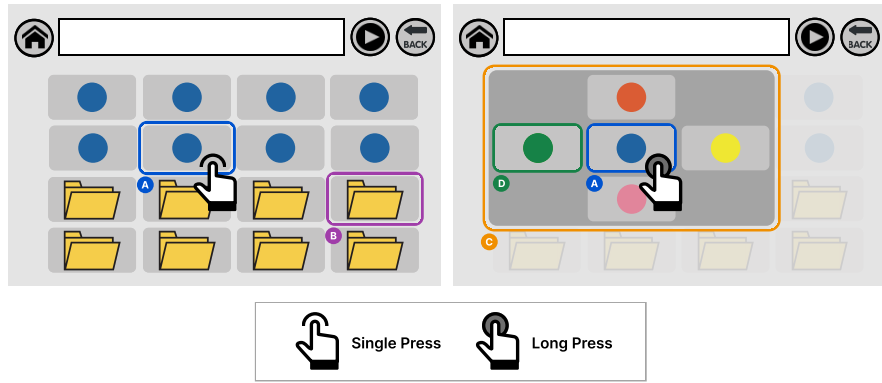}
\caption{On the left, a traditional grid display layout is described with communication options (A) and folder buttons for sub-page navigation (B). On the right, our design for predictive anchoring is shown. An overlay containing a radial display (C) contains the initial communication option the user is interested in (A) with predictions for other vocabulary they may be interested in (D). In our design on the right, the single press interaction modality of the traditional grid display is preserved, with the optional predictions provided by the long press interaction.}
\label{fig:design}
\Description[An image depicting the proposed design.]{On the left hand side, a traditional grid display layout is depicted with two points of interaction highlighted. The first point of interaction highlighted is a vocabulary button, represented by a circle. The second point of interaction highlighted is a folder button, represented by a folder icon. Both points of interaction are completed with a single press. On the right hand side, an abstracted version of the proposed design of anchoring communication options is depicted. The traditional grid layout has been overlaid with a window which contains a radial layout. This overlay is toggled via a long press. Users can still use the single press to use the AAC device in the traditional manner. The vocabulary word the user indicated that they are interested is in the center, represented by the same circles from the traditional grid display. Surrounding the initial communication option are four recommendations which are represented by four different colored circles (green, orange, yellow, and pink).}
\end{figure*}

In this work, we propose an investigation into a new design interaction for grid displays which leverage an anchoring technique to enhance grid display communication (see Figure~\ref{fig:design}). 
Our design preserves the existing layout of grid displays (see Design Goal 2). A user of this grid display would be able to use it in the widely accepted manner, selecting vocabulary and navigating through folders and sub-folders in order to locate and select vocabulary. Users would also be able to select a communication option for the enhanced version of the grid display via a long-press. If the user long presses on a communication option an overlay will appear with suggestions for other vocabulary they may be interested in as depicted in Figure~\ref{fig:design} (see Design Goal 1). For example, if the user long-presses on ``eat'', an overlay with the following suggestions of their favorite foods would appear. Additionally, this interaction would extend to folders. If a folder was long-pressed, suggestions from within the underlying folder structure would be suggested, in order to preserve mental models (see Design Goal 2). These suggestions are intended to both enhance the rate of communication of users of grid displays (see Design Goal 1) and anchor vocabulary together to aid in linguistic development and provide more accessible vocabulary (see Design Goal 3).

\section{Discussion \& Future Work}
In this work, we have proposed a novel interaction technique meant to enhance existing grid display interfaces for beginning communicators. Grid displays possess a number of benefits, especially for communicators who require access to a large amount of vocabulary~\cite{gevarter2020dynamic, light_performance_2004}; however, the layout of grid displays can pose a number of difficulties for less experienced users and communicators with limited vocabulary~\cite{light2019new, light2007aac, shane2006using}. Additionally, the lack of context and anchoring provided in grid displays makes developing language more complex for beginning communicators~\cite{light1997let}. In spite of this, grid displays are still one of the most common forms of AAC~\cite{thistle2015building}. We have presented an interaction technique which takes advantage of users' existing grid display interfaces by optionally providing contextually-relevant vocabulary to enhance the rate of communication of users and benefit linguistic development.

As future work, we need to evaluate this system with communication partners and end users of AAC to evaluate this system design and ensure that this design and interaction pattern serves users in the most effective way. 
Future work must also critically analyze the potential impacts of predictions, both correct and incorrect, and potential biases and harm from AI systems~\cite{mehrabi2021survey} on end users and communication partners. Additionally, there are open questions regarding what the correct method of interaction with the predictive anchoring overlay is and how many options should be presented. 

\begin{acks}
This work was partially funded by the Temple University Pervasive Computing REU Site and Convergence Accelerator Grant (National Science Foundation grant numbers CNS-2150152 and ITE-2236352). 
\end{acks}

\bibliographystyle{ACM-Reference-Format}
\bibliography{sample-base}

\end{document}